\documentclass[12pt]{article}

\usepackage{amsmath}    
\usepackage{amssymb}
\usepackage{bbm}
\usepackage{subeqnarray}
\newcommand{\vrh}{\hbox{$\mathbb{H}$}}
\newcommand{\vrk}{\hbox{$\mathbb{K}$}}
\newcommand{\vrg}{\hbox{$\mathbb{G}$}}
\newcommand{\vrq}{\hbox{$\mathbb{Q}$}}
\newcommand{\vrz}{\hbox{$\mathbb{Z}$}}
\newcommand{\vrjed}{\hbox{$\mathbbm{1}$}}
\begin{document}
\def\half{{\textstyle\frac{1}{2}}}
%

\title{D=4 Extended Galilei Superalgebras with Central Charges}

\author{JERZY LUKIERSKI
\\[14pt]
Institute for Theoretical Physics\\
University of Wroc{\l}aw, pl. Maxa Borna 8,
\\
50-204 Wroc{\l}aw
Poland \\
lukier@ift.uni.wroc.pl}

\date{}
\maketitle

\begin{abstract}
We perform a nonrelativistic contraction of $N$-extended Poincar\'{e} superalgebra with internal symmetry $U(N)$ and general set of $N(N-1)$ real central charges. We show that for even $N=2k$ and particular choice of the dependence of $Z_{ij}$ on light velocity $c$ one gets the $N$-extended Galilei superalgebra with unchanged number of central charges and compact internal symmetry algebra 
$U(k;H)=USp(2k)$.
 The Hamiltonian positivity condition is modified only by
 one central charge.
If we put all the central charges equal to zero one gets the $2k$-extended Galilei superalgebra as the subalgebra of recently introduced extended Galilei conformal superalgebra \cite{phluk1,phluk2}.
\end{abstract}

\section{Introduction}	
The Galilean symmetry is the fundamental geometric property of the nonrelativistic systems. For the description of quantized massive $d$-dimensional nonrelativistic systems one uses the Galilean projective representations (see e.g. \cite{phluk3}), obtained in the presence of central extension of $D=d+1$ Galilei algebra. Such centrally extended Galilei algebra is called quantum Galilei or Bargmann algebra. The case $D=2+1$ is a special one, it permits to add second ``exotic'' central charge $\theta$ \cite{phluk4,phluk5}, introducing the noncommutativity of two-dimensional Galilean boosts generators\footnote{In \protect\cite{phluk6} it was shown that the second central charge $\theta$ can be interpreted as describing the noncommutativity of $d=2$ space coordinates.}.

In this paper we derive the most general supersymmetric extension of $D=3+1$ Galilei algebra. We recall that $d=3$ Galilei algebra has the form ($i,j,k=1,2,3$)

\begin{eqnarray}\label{pheq1}
&[P_i, B_j ] = i\, \delta_{ij} m 
\qquad [P_i, P_j ] = [ B_i , B_j] = 0&
\cr\cr
&[I_i , I_j ] = i\, \varepsilon_{ijk} \, I_k&
\cr\cr
&[I_i , P_j ] = i\, \varepsilon_{ijk} \, P_k
\qquad
[I_i , B_j ] = i\, \varepsilon_{ijk} \, B_k&
\cr\cr
&[H,B_i] = i\, P_i 
\qquad \quad
[H, I_i] = [H, P_i ] =0&\,,
\end{eqnarray}
where $P_i$ describe momenta, $B_i$ - Galilean boosts,
 $H$ - the energy operator (Hamiltonian) and $I_i$ generate $O(3)$ rotations. It is well-known that the Galilei algebra (\ref{pheq1}) can be obtained from $D=4$ Poincar\'{e} algebra ($P_\mu=(P_i, P_0)$, $M_{\mu\nu} =(M_i, N_i)$) by the relativistic contraction $c\to\infty$
  \cite{phluk7} if we perform the following $c$-dependent rescaling ($P_i, M_i$ remain unchanged)
  \begin{equation}\label{pheq2}
  N_i = c\, B_i \qquad \qquad P_0 = c\, m_0 + \frac{H}{c}\,.
  \end{equation}
  
  In this paper we shall extend such a contraction procedure to the $N$-extended $D=4$ superPoincar\'{e} algebra with maximal number of $N(N-1)$ real central charges \cite{phluk8,phluk9}. 
  In derived $N$-extended Galilei superalgebra 
  the generators ($H, P_i, B_i$) and the $N(N-1)$ real central charges
  can be expressed as bilinears of supercharges.
  We add that such superalgebra we could derive 
  our superalgebra  only for even $N$ ($N=2k$).
  
  The first attempt of supersymmetrizing the  $D=3+1$ Galilei algebra was made by Puzalowski \cite{phluk10}, who considered $N=1$ and $N=2$ cases. We recall that already in seventies the Galilean symmetries were extended to so-called Schr\"{o}dinger symmetries of free quantum-mechanical systems \cite{phluk11,phluk12} by
  supplementing Galilei algebra with two additional generators: $D$ (dilatations) and $K$ (extensions, describing conformal time transformations). The Schr\"{o}dinger  algebra was then supersymmetrized (see e.g. \cite{phluk13}--\cite{phluk19}) and
   in such a case the Galilei superalgebra can be obtained as the sub-superalgebra.
  However in such a way the $N$-extended supersymmetrization of $D=3+1$ Galilei algebra has not been described in its general form, because the presence of additional generators $D,K$ in Schr\"{o}dinger  superalgebra does not permit to introduce all central charges, allowed by SUSY extension of smaller Galilei algebra. The general consideration of the supersymmetrization of Galilei algebras has been presented only for ``exotic'' $D=2+1$ case in \cite{phluk20}. In this paper we shall firstly obtain the complete results in the ''physical'' case $D=3+1$.
  
  Firstly in Sect.~2 we shall recall, using real Majorana representation, the $N$-extended $D=4$ Poincar\'{e} algebra as expressed in terms of projected supercharges. 
  In Sect.~3 we introduce for even $N$ ($N=2k$) the projected supercharges,  which before the contraction are rescaled by 
  powers of $c$ in homogeneous way.
  In Sect.~4 we shall perform the nonrelativistic contraction limit  $c\to\infty$
   providing the supersymmetrization of seven Galilean generators, $H, P_i, B_i$ and all central charges.  We show as well how in the presence of arbitrary Galilean central charges the Hamiltonian positivity condition is modified.
   In final Sect.~5 we consider the link with recently derived Galilei conformal supalgebra \cite{phluk1,phluk2}, simplification of central charges sector by changing
    by similarity transformation  the $U(N)$ basis of supercharges, and we provide remarks about the Galilean tensorial central charges.
    We stress that our calculations can be performed as well in
    complex Weyl supercharges basis, and we plan to use it in
    the future discussion of the realizations of Galilei
    superalgebra with central charges.
  
  We add that recently there were considered the nonrelativistic super-$p$-branes \cite{phluk21,phluk22} and supersymmetric $D$-branes \cite{phluk23}. Following the relativistic considerations (see e.g. \cite{phluk24}) 
   one can conjecture that the nonrelativistic central charged should be associated with topological defects, e.g. with vortices (superstrings) or domain walls (supermembranes).
   
   \section{Relativistic D=4 Extended Poincar\'{e} Superalgebras with Central Charges}
   
   The general $N$-extended D=4 Poincar\'{e} superalgebra in real Majorana formulation is described by
   \\[14pt]  
   {\bf i)}   
   $4N$ real supercharges $Q^r_A$ ($r=1,\ldots N,\  A=1 \ldots 4$)
\\[14pt]   
{\bf ii)}
   Poincar\'{e} algebra generators $\ P_\mu =(P_0,P_i)$,
    $\qquad M_{\mu\nu}=(M_i=\half \varepsilon_{ijk} M_{jk}$,  $ \qquad  N_i=M_{i0})$
    \\[14pt]  
    {\bf iii)}
    Internal $U(N)$ symmetry generators represented by a pair of symmetric ($T^{rs}_{(s)}=T^{sr}_{(s)}$) and antisymmetric 
    ($T^{rs}_{(a)}= -T^{sr}_{(a)}$) sets of anti-Hermitean generators, satisfying the relations
    \begin{equation}\label{pheq3}
    \begin{array}{l}
    [T^{sr}_{(S)}, T^{tu}_{(S)}]
    = \delta^{rt} T^{su}_{(A)} + \delta^{st} T^{ru}_{(A)}
    + \delta^{su} T^{rt}_{(A)} + \delta^{ru} T^{st}_{(A)}
    \cr\cr
    [T^{sr}_{(S)}, T^{tu}_{(A)}]
    = \delta^{su} T^{rt}_{(S)} + \delta^{ru} T^{st}_{(S)}
    - \delta^{st} T^{ru}_{(S)} - \delta^{rt} T^{su}_{(S)}
    \cr\cr
    [T^{sr}_{(A)}, T^{tu}_{(A)}]
    = \delta^{rt} T^{su}_{(A)} - \delta^{st} T^{ru}_{(A)}
    + \delta^{su} T^{rt}_{(A)} - \delta^{ru} T^{st}_{(A)}
    \, .
    \end{array}
    \end{equation}
    One can check that the axial generator $A=T^{rr}_{(S)}$ commutes with $T^{rs}_{(S)}, T^{rs}_{(A)}$ and can be separated out 
    ($U(N)=SU(N)\oplus U(1)$); the traceless generators 
    $\widetilde{T}^{rs}_{(S)}=T^{rs} - \frac{1}{N}
    \delta^{rs}A$  and $T^{rs}_{(A)}$ describe then the internal symmetry $SU(N)$.
    
    {\bf iv)} The set of $N(N-1)$ real central charges 
    $X^{rs} = - X^{sr}, Y^{rs} = - Y^{sr}$.
    
    The basic $N$-extended SUSY relation has the following form (see e.g. \cite{phluk9})
    \begin{eqnarray}\label{pheq4}
    \{ Q^{r}_{A}, Q^{s}_B \} = \delta^{rs} (\gamma^\mu C)_{AB} \, P_\mu 
    + X^{rs}\, C_{AB} + Y^{rs}(\gamma_5\, C)_{AB}\,,
    \end{eqnarray}
    where we use real Majorana realization of Dirac matrices satisfying the relations
    \begin{equation}\label{pheq5}
    \gamma_i=\gamma^T_i \qquad
    C= \gamma_0= - \gamma^T_0 \qquad
    \gamma_5=- \gamma^T_5 \, .
    \end{equation}
    The bosonic-fermionic sector of the superalgebra (covariance relations) looks as follows:

\begin{subequations}\label{pheq6}
\begin{align}   
    \label{pheq6a}
  & [P_\mu, Q^r_A ] =0
  \nonumber  \\[10pt]
    &
    [M_{\mu\nu} , Q^r_{A}] =
    - \half (\sigma_{\mu\nu})_{AB}\, Q^r_{B}
     \\ \cr
    \label{pheq6b}
    &
    [T^{rs}_{(S)} , Q^t_{A} ]
     = (\gamma_5)_{A}^{\ B}
    (\tau^{\ \  rs}_{(S)})^t_{\ u}\,Q^u_B 
    \nonumber  \\[10pt]
    &
    [T^{rs}_{(A)} , Q^t_{A} ]= 
    (\tau^{\ \ rs}_{(A)} )^t_{\ u}\,Q^u_A
    \\ \cr
    \label{pheq6c}
  &
    [X^{rs} , Q^t_{A} ] = [Y^{rs}, Q^t_A ]=0 \,,
    \end{align}
   \end{subequations}

\noindent
where the $N\times N$ matrices $\tau^{rs}_{(S)}, \tau^{rs}_{(A)}$ are the fundamental matrix realizations of algebra (\ref{pheq3}).

The Abelian charges $X^{rs}, Y^{rs}$ form the tensorial $(2,0)$ representation of internal symmetry $U(N)$. Introducing complex generators
\begin{equation}\label{phequ7}
T^{rs} = T^{rs}_{(S)} + i \, T^{rs}_{(A)}
\qquad
Z^{rs} = X^{rs} + i \, Y^{rs} \,,
\end{equation}
one gets the standard covariance relations for the holomorphic $(2,0)-U(N)$ tensors
\begin{equation}\label{pheq8}
[ T^{rs} , Z^{tu} ]  
= \delta^{rt} \, Z^{su} 
+ 
\delta^{st} \, Z^{ru}
-
\delta^{ru} \, Z^{st}
-
\delta^{su} \, Z^{rt}\, .
\end{equation}

If we assume that the central charges $Z^{rs}$ are described by a numerical matrix $Z^{rs}_{(0)}$, their presence breaks internal $U(N)$ symmetry. For a definite configuration of numerical values of $Z^{rs}_{(0)}$ one can determine the residual internal symmetry with the generators commuting with the constant  matrix of central charges. 
 (see e.g. \cite{phluk25}).
    
\section{Projection Procedure and New Form of N-extended D=4 Poincar\'{e} Algebra}

Before performing the nonrelativistic contraction $c\to \infty$ we shall introduce the projected supercharges. We shall use the following projection operators (see also \cite{phluk1})
\begin{equation}\label{pheq9}
P^{rs}_{\pm AB} = \half \left(\delta^{rs}\, \delta_{AB}
+ \Omega^{rs}\, C_{\alpha \beta}\right)\, ,
\end{equation}    
where $\Omega^{rs}= - \Omega^{sr}$ and $\Omega^2 = -1$. Because $C^2=-1$ we get the required projector properties
\begin{equation}\label{pheq10}
P^{rs}_{\pm AB} \, P^{st}_{\pm BC}  = 
P^{rt}_{\pm AC} 
\qquad \quad
P^{rs}_{\pm AB} \, P^{st}_{\mp BC} = 0 \, .
\end{equation}
If we define projected supercharges as follows ($r,s=1 \ldots  N=2k$)
\begin{equation}\label{pheq11}
Q^{ \ r}_{\pm A} = 
P^{rs}_{\pm AB}  \, Q^s_B \, ,
\end{equation}
from (\ref{pheq4}) we get

\begin{subequations}
\begin{align} 
\label{pheq12a}
&
\left\{
Q^{ \ r}_{\pm A}, Q^{ \ s}_{\pm B}
\right\}
= 
P^{rs}_{\pm AB} \, P_0 
+
P^{rt}_{\pm AC} 
\left(
C_{CB}\, X^{ts}_{+} + (\gamma_5 \, C)_{CB}
\, Y^{ts}_{-} 
\right)
\phantom{xxxxxxssx}
\\
\cr
\label{pheq12b}
&
\left\{
Q^{ \ r}_{+ A}, Q^{ \ s}_{- B}
\right\}
= 
P^{rs}_{+ AC} 
(\gamma^i\, C)_{CB} \, P_i
+
P^{rt}_{+ AC}
\left(
C_{CB}\, X^{ts}_{-} 
+ (\gamma_5 \, C)_{CB} \, Y^{ts}_{+} 
\right) 
\end{align}
   \end{subequations}
where
\begin{equation}\label{pheq13}
X_{\pm}\, \Omega = \pm \Omega\, X_{\pm}
\qquad
Y_{\pm}\, \Omega = \pm \Omega\, Y_{\pm}\, ,
\end{equation}
or more explicitly (for $Y_{\pm}$ the formulae are analogous)
\begin{equation}\label{pheq14}
X_{+} = \begin{pmatrix} A & B \cr -B &A
\end{pmatrix}
\qquad \quad
X_{-} = \begin{pmatrix} A & \widetilde{B} \cr 
\widetilde{B} & -A
\end{pmatrix}
\end{equation}
where $A= -A^T$, $B=B^T$, $\widetilde{B} = - \widetilde{B}^T$ are $k\times k$ matrix blocks and 
${\rm dim} \, X_{+} = k^2, \ {\rm dim} \, X_{-} = k(k-1)$.

The nonvanishing commutators with projected supercharges following from relations (\ref{pheq6}) are

\begin{equation}\label{pheq15}
\begin{array}{lll}
 [ M_{\mu\nu}, Q^{\ r}_{\pm A} ]
& =  &
- \half (\sigma_{\mu\nu})_{A}^{\ B}\, Q^{\ r}_{\pm B}
 \\[18pt]
 [ T^{rs}_{+ (S)}, Q^{\ t}_{\pm A} ]
 & = &
(\gamma_5)_{A}^{ \ B}
(\tau_{+ (S)}^{ \quad rs})^{t}_{\ u}
\, Q^{\ u}_{\mp B}
 \\[18pt]
 [ T_{- (S)}^{rs}, Q^{\ t}_{\pm A} ]
 & = &
(\gamma_5)_{A}^{ \ B}
(\tau_{- (S)}^{ \quad rs})^{t}_{\ u}
\, Q^{\ u}_{\pm B}
 \\[18pt]
 [ T_{+ (A)}, Q^{\ t}_{\pm A} ]
 & = &
 (\tau_{+ (A)}^{\quad rs})^{t}_{\ u}\, Q^{\ u}_{\pm A}
 \\[18pt]
 [ T_{- (A)}, Q^{\ t}_{\pm A} ]
 & = &
 (\tau_{- (A)}^{\quad rs})^{t}_{\ u}\, Q^{\ u}_{\mp A} \, ,
\end{array}
\end{equation}
\\[10pt]
where $ T_{ (S)}^{rs} =  T_{+ (S)}^{rs} +
 T_{- (S)}^{rs}$, 
 $ T_{ (A)}^{rs} =  T_{- (A)}^{rs} +
 T_{+ (A)}^{rs}$, and
 \begin{equation}\label{pheq16}
 T_{\pm (S)} \, \Omega = \pm \Omega \, T_{\pm (S)}
 \qquad 
 T_{\pm (A)} \, \Omega = \pm \Omega \, T_{\pm (A)} \, .
 \end{equation}
The relations (\protect\ref{pheq16}) imply that
\begin{subequations}
\begin{align}
\label{pheq17a}
\Omega \cdot \vrh {}
= - \vrh{}^T \, \Omega
\qquad \quad
\vrh{} = (T_{- (S)}, T_{+ (A)} )
\\
\cr
\label{pheq17b}
\Omega \cdot \vrk{}
= - \vrk{}^T \, \Omega
\qquad \quad
\vrk{} = (T_{+ (S)}, T_{- (A)} ) \, .
\end{align}
\end{subequations}
The symplectic condition (\ref{pheq17a}) restricts the $N\times N$ ($N=2k$) matrix realizations of the $U(2k)$ generators to the subalgebra $USp(2k)=U(2k)\cap Sp(2k)$, i.e.
\begin{equation}\label{pheq18}
\vrh{} = {\rm usp}(2k) \simeq U(k;H)
\qquad \quad 
\vrk{} = \frac{U(2k)}{U(k;H)} \, .
\end{equation}
The pair (\ref{pheq18}) describes the symmetric Riemannian pair $(\vrh{},\vrk{})$ satisfying the relations
\begin{equation}\label{pheq19}
[\vrh{}, \vrh{} ] \subset \vrh{}
\qquad
[\vrh{}, \vrk{} ] \subset \vrk{}
\qquad
[\vrk{}, \vrk{} ] \subset \vrh{} \, .
\end{equation}
We obtain as well ($Q^{r}_{+A} \subset \vrq{}_{\pm }$)
\begin{eqnarray}\label{pheq20}
&&[ \vrh{}, \vrq_{\pm} ] \subset \vrq_{\pm}
\cr\cr
&&[\vrk{}, \vrq{}_{\pm} ] \subset \vrq{}_{\mp} \,.
\end{eqnarray}
The $2k(2k-1)$ central charges $\vrz{} = (X^{rs}, Y^{rs})$
 are split into the following two sets
 \begin{equation}\label{pheq21}
 \vrz{}_1 = (X^{ts}_{+} , Y^{ts}_{-} )
 \qquad \quad
 \vrz{}_2 = (X^{ts}_{-}, Y^{ts}_{+}) \,,
 \end{equation}
 both with equal dimensionality $2(2k-1)$. 
 We see from (\ref{pheq12a}--\ref{pheq12b}) that the central charges $\vrz{}_1$ occur in the anticommutator
 $\{ Q_{\pm}, Q_{\pm}\}$, and $\vrz{}_{2}$ in the 
 anticommutator $\{ Q_{\pm}, Q_{\mp}\}$.
 
 The supercharges (\ref{pheq11}) due to the relation (\ref{pheq10}) satisfy the constraints
 \begin{equation}\label{pheq22}
 P^{rs}_{\mp AB} \, Q^{\quad s}_{\pm B} = 0 \,.
 \end{equation}
 Choosing in (\ref{pheq9}--\ref{pheq11})
 \begin{equation}\label{pheq23}
 \Omega = \begin{pmatrix}
 0 & {\boldsymbol\vrjed{}}_k
 \cr - {\boldsymbol\vrjed{}}_k & 0
 \end{pmatrix}
 \end{equation}
one can express (\ref{pheq22}) as the following relation between the $2k$ components of $Q^{\ r}_{\pm}$:
\begin{equation}\label{pheq24}
Q^{k+i}_{\pm A}
= \mp (\gamma_0)_{AB} \, Q^{i}_{\pm B}
\qquad \quad (i=1\ldots k)\, ,
\end{equation}
where we inserted $C=\gamma_0$. We see therefore that in the algebraic relations (\ref{pheq12a}--\ref{pheq12b}) the independent components are described by indices $r,s=1,2 \ldots k$.

\section{Nonrelativistic Contraction Procedure and the Extended Galilei Superalgebras}

In previous section we prepared the algebraic framework convenient for obtaining the finite nonrelativistic $c\to \infty$ limit.

Let us supplement the rescalings (\ref{pheq2}) by the following ones (see also \cite{phlukxx})
\begin{equation}\label{pheq25}
Q^{\ r}_{+ A} = \frac{1}{\sqrt{c}} \, \widetilde{Q}^{\, r}_{A} \qquad \quad 
Q^{\ r}_{- A} = {\sqrt{c}} \, \widetilde{S}^{\, r}_{A} \,.
\end{equation}
Introducing the split
\begin{equation}\label{pheq26}
X^{rs}_{+} = Z \, \Omega^{rs} + \widetilde{X}^{\, rs}_{+}
\end{equation}
we postulate further
\begin{eqnarray}\label{pheq27}
&&
Z = - \frac{1}{2} \, m_0 \, c + 
\frac{1}{2c}\, U + O(\frac{1}{c^2})
\cr\cr
&&
\widetilde{X}^{rs}_{+} = \frac{1}{c}\, U^{rs}_{+}
+ O(\frac{1}{c^2})
\cr\cr
&&
Y^{rs}_{-} = \frac{1}{c} \, U^{rs}_{-} + O(\frac{1}{c^2})
\end{eqnarray}
and assume that $(X^{rs}_{-}, Y^{rs}_{+})$ have the finite
$c\to \infty$ limits $(V^{rs}_{-}, V^{rs}_{+})$. Using the identities
\begin{equation}\label{pheq28}
P^{rt}_{\pm AC} \, C_{CB} \, \Omega^{ts}
=
\pm \, 2 \,  P^{rs}_{\pm AB}
\end{equation}
we obtain from (\ref{pheq12a},\ref{pheq12b})
in the limit $c\to \infty $ the following basic Galilei superalgebra relations
\begin{subequations}\label{pheq29}
\begin{align}
\label{pheq29a}
&
\left\{
\widetilde{Q}^{r} _{A}, \widetilde{Q}^{s}_{B}
\right\}
= P^{rs}_{+ AB}
( H+U)
+
P^{rt}_{+ AC}
\left(
C_{CB} \,U^{rt}_{+} +
(\gamma_5 \, C)_{CB} \, U^{rt}_{-}
\right)
\\
\label{pheq29b}
\cr
&
\left\{
\widetilde{Q}^{r} _{A}, \widetilde{S}^{s}_{B}
\right\}
= P^{rs}_{+ AC}
(\gamma^{i}\, C)_{CB} \, B_i
+
P^{rt}_{+ AC}
\left(
C_{CB} \, V^{ts}_{-} +
(\gamma_5 \, C)_{CB} \, V^{ts}_{+}
\right)
\\ \label{pheq29c}
\cr
&
\left\{
\widetilde{S}^{r} _{A}, \widetilde{S}^{s}_{B}
\right\}
= P^{rs}_{- AB} \cdot 2 \, m_0 \, .
\end{align}
\end{subequations}

The consistent $c\to \infty$ limit of (\ref{pheq15}) requires that the generators (\ref{pheq18}) are rescaled 
 as follows:
 \begin{equation}\label{pheq30}
 \vrh{} = \widetilde{\vrh{}} 
 \qquad \quad
 \vrk{} = c \, \widetilde{\vrk{}} \, .
 \end{equation}
 One obtains

\begin{subeqnarray}\label{pheq31}
\label{pheq31a}
&&
[M_i, \widetilde{Q}^{r}_{A} ]
= 
- \half (\sigma_i)_{A}^{\ B} \,
\widetilde{Q}^{r}_{B}
\cr\cr
&&
[M_i, \widetilde{S}^{\, r}_{A} ]
 = 
- \half (\sigma_i)_{A}^{\ B} \,
\widetilde{S}^{\, r}_{B}
\\[16pt]
\label{pheq31b}
\cr
&&
[B_i, \widetilde{Q}^{r}_{A} ]
 =
[B_i, \widetilde{S}^{\, r}_{A} ] =0
\\[16pt]
\label{pheq31c}
\cr
&&
[\widetilde{T}^{rs}_{-(S)}, \widetilde{Q}^{t}_{A} ]
 =
(\gamma_5)^{\ B}_{A} (\tau^{\quad rs}_{-(S)})^{\, t}_{\ u}
\, \widetilde{Q}^{u}_{B}
\cr
\cr
&&
[\widetilde{T}^{rs}_{-(S)}, \widetilde{S}^{\, t}_{A} ]
 =
(\gamma_5)^{\ B}_{A} (\tau^{\quad rs}_{-(S)})^{\, t}_{\ u}
\, \widetilde{S}^{\, u}_{B}
\cr
\cr
&&
[\widetilde{T}^{rs}_{+(A)}, \widetilde{Q}^{t}_{A} ]
 =
(\tau_ {+(A)}^{\quad rs})^{\, t}_{\ u}\, \widetilde{Q}^{u}_{A}
\cr
\cr
&&
[\widetilde{T}^{rs}_{+(A)}, \widetilde{S}^{\, t}_{A} ]
 =
(\tau_ {+(A)}^{\quad rs})^{\, t}_{\ u}\, \widetilde{S}^{\,u}_{A}
\\[16pt]
\label{pheq31d}
\cr
&&
[\widetilde{T}^{rs}_{+(S)}, \widetilde{Q}^{t}_{A} ]
 =
[\widetilde{T}^{rs}_{+(S)}, \widetilde{S}^{\, t}_{A} ]
= 0
\cr
\cr
&&
[\widetilde{T}^{rs}_{-(A)}, \widetilde{Q}^{t}_{A} ]
 =
[\widetilde{T}^{rs}_{-(A)}, \widetilde{S}^{t}_{A}]=0 \, .
\end{subeqnarray}
Using the relations (\ref{pheq19}) we obtain after the contraction $c\to \infty$ that
\begin{equation}\label{pheq32}
\widetilde{\vrh{}} \in USp(2k) \simeq U(k,H)
\end{equation}
and the generators $\widetilde{\vrk}$ become Abelian, transforming as the linear representation of the internal symmetry algebra $\widetilde{\vrh{}}$.

Because $[\widetilde{\vrh{}}, \widetilde{\vrk{}}]\sim \widetilde{\vrk{}}$ are the only nonvanishing commutators of the contracted generators $\widetilde{\vrk{}}$, one can consistently remove these generators from the Galilei superalgebra $\vrg{}^{(2k)}$, which can be described as the semidirect product
\begin{equation}\label{pheq33}
\vrg{}^{(2k)} = (O(3)\oplus U(k;H))
\ltimes
{\widetilde{\cal A}}
\end{equation}
where $\widetilde{\cal A}$ is the superalgebra described by the relations (\ref{pheq29a}--\ref{pheq29c}). Using
 the consequences of the formulae (\ref{pheq9}), (\ref{pheq13}) and  (\ref{pheq14})
\begin{equation}\label{pheq34}
{\rm tr} \left(
P^{rs}_{\pm AB}
\right) = 4N
\qquad
{\rm tr}(\Omega\, U_+) =0
\end{equation}
one obtains that
\begin{equation}\label{pheq35}
H + U = 
\frac{1}{4N} \, {\widetilde{Q}}^{r}_{A}\, 
{\widetilde{Q}}^{r}_{A} \geq 0
\end{equation}
and
\begin{equation}\label{pheq36}
m_0 = \frac{1}{8N} \, 
{\widetilde{S}}^{\, r}_{A} \, 
{\widetilde{S}}^{\, r}_{A} \geq 0 \,.
\end{equation}
The relation (\ref{pheq35}) describes the generalization of the energy positivity condition in the presence of general central charges. 
\section{Final Remarks}

The aim of this paper is to present the most general extended Galilei superalgebra in D=4. The supersymmetrization is obtained by nonrelativistic contraction of $N$-extended Poincar\'{e} superalgebra for even $N=2k$. The necessity of even $N$ can be explained if we observe that the compact Galilean internal symmetry is described by quaternionic algebra $U(k;H)=USp(2k)$ which requires even number of complex supercharges. Also it should be mentioned that for $N=2k$ we are able to supersymmetrize (i.e. express as bilinear expression in terms of supercharges) the seven-dimensional Abelian subalgebra ($H,P_i, B_i$) of D=4 Galilei algebra (see (\ref{pheq1})) as well as Abelian central charges. We conjecture that such full supersymmetrization requires $N=2k$.~{}\footnote{An argument in this favour is provided by an effort of obtaining $N=1$ supersymmetrization of Galilei conformal algebra \protect\cite{phluk24} which does not lead to Hamiltonian as bilinear form in terms of supercharges.}

In relation with presented algebraic structure of D=4 $2k$-extended Galilei superalgebra we have the following comments:

{\bf i)} If we put all central charges equal to zero it can be checked that the extended Galilei superalgebra can be obtained as the sub-superalgebra of extended Galilei conformal superalgebra \cite{phluk1,phluk2}. Such sub-superalgebra can be obtained by considering only half of Galilean conformal supercharges, and putting the generators $D$ (dilatations), $K$ (expansions) and $F_i$ (Galilean conformal translations describing constant accelerations)  equal to zero. It can be recalled that in both Galilei and Galilei conformal superalgebras the internal compact symmetries are described by the algebra $U(k;H)$. 

{\bf ii)} An interesting task is to consider the Hilbert space realizations of extended Galilei superalgebras with central charges. Before this task one can consider the unitary transformations $U(k)$  putting the antisymmetric Galilean central charges 
$U^{rs} = - U^{sr}$, $V^{rs} = - V^{sr}$ in Jordanian form
\begin{equation}\label{pheq37}
\begin{array}{l}
U^{rs} = (i \, \sigma_2) \otimes 
{\rm diag}\, (u_1,\ldots u_k)
\cr\cr
V^{rs} =
(i\, \sigma_2) \otimes {\rm diag} (v_1 \ldots v_k)
\end{array}
\end{equation}

For studying the realizations it is more convenient to consider the superalgebra (\ref{pheq29a}--\ref{pheq29c}) written in terms of two-component complex (Weyl) supercharges\footnote{In Weyl representation the $\gamma_5$-matrix is replaced by complex unit $i$, i.e. one gets $\widetilde{Q}^{\, r}_{\pm A} +
\gamma_5\, \widetilde{S}^{\, r}_{\pm A} \to 
\widetilde{Q}_{\pm \alpha} + i \widetilde{S}^{\, r}_{\pm \alpha}$ ($\alpha=1,2$) and 
$U^{rs} + \gamma^5\, V^{rs} \to
Z^{rs} = U^{rs} + i\, V^{rs}$.}. It is interesting, following the discussion of relativistic case (see e.g. \cite{phluk25}), to consider the structure of nonrelativistic particle states in the presence of central charges e.g. if $N=4$ ($k=2$) and $N=8$ ($k=4$).

{\bf iii)} We restricted our considerations to the case with scalar central charges \cite{phluk8}, however recently, following the importance of $D=11$ $M$-algebra, there were considered generalized D=4 superalgebras with tensorial central  charges (see e.g. \cite{phluk26,phluk27}). These generalized space-time algebras should have also the nonrelativistic $c\to \infty$ limit related with nonrelativistic superstrings \cite{phluk21,phluk22} and supermembranes \cite{phluk23}.

\subsection*{Acknowledgments}
The author would like to thank P. Horv\'athy,  E. Ivanov 
and P. Kosi\'{n}ski for discussion and Polish Ministry of Science and Higher Education (projects NN202318534 and NN202331139) for financial support.

\end{document}